\documentclass[aps,showkeys]{revtex4}

\usepackage{epsfig}
\usepackage{epsf}

\def \be{\begin{equation}}
\def \ee{\end{equation}}

\begin{document}

\title{Finding the Dominant Zero of the Energy Probability Distribution}%


\author{J. J. Carvalho}
\email{jaderjcarvalho@ufsj.edu.br}
\author{A. L. Mota}
\email{motaal@ufsj.edu.br}
\affiliation{Departamento de Ci\^{e}ncias Naturais, Universidade Federal de S\~{a}o Jo\~{a}o del Rei,
C.P. 110,  CEP 36301-160, S\~ao Jo\~ao del Rei, Brazil}

\begin{abstract}
In this work, we present a method to locate the dominant zero of the Energy Probability Distribution (EPD) Zeros method applied to the study of phase transitions. The technique strongly reduces computer processing time and also increases the accuracy of the results. As an example, we apply it to the 2D Ising model, comparing with both the exact Onsager's results and the previous EPD zeros method results. We show that, for large lattices, the processing time is drastically reduced when compared to other EPD zeros search procedures, whereas for small lattices the gain in accuracy allows very accurate predictions in the thermodynamic limit.

\end{abstract}

\maketitle

\section{Introduction}

Computer simulation and finite-size scaling theory, together, constitute one of the most relevant tools in the study of phase transitions and critical phenomena. Since exact solutions of statistical mechanics models for condensed matter systems are available only for a few cases, and other analytical approaches always involve some level of approximation, accuracy can be obtained, in general, only through simulation, as long as the necessary computational time is feasible and available. However, as the number of available statistical states grows (usually with the system dimensions), as well as the models' complexity also increases, more computer time is necessary to reach some required precision. Thus, efficient techniques to locate the critical points, to reduce the computer processing time, and to increase the accuracy of the results are very desirable.

One of the most recent techniques that fulfill these requirements is the Energy Probability Distribution (EPD) method \cite{Rocha2016,bvcosta2017}. It is a new, general and efficient method that allows the obtaining of the transition temperature and critical exponents for a given statistical mechanics system with high precision in a broad range of applications \cite{Figueiredo2017,bvcosta20172,Mol2018,bvcosta2019,bvcosta20192,Mol2019}. It can be also very accurate, depending, of course, on the accuracy of the energy probability distribution employed. It is based on the Fisher zeros of the partition function \cite{Fisher1965}, an issue that is, recently, getting more attention in the study of critical phenomena \cite{Kim2017,Deger2019}. The EPD zeros method can be summarized as follows.

The partition function of a thermodynamic system is given by 
\be
Z=\sum_E g(E) e^{-\beta E}, 
\ee
where $g(E)$ is the number of states with a given energy $E$ and $\beta=1/T$ is the inverse temperature, in units of the inverse Boltzmann constant, $k_B^{-1}$. Assuming that the allowed energies can be casted in a discrete sequence 
\be
E_n = E_0 + n \delta E,
\ee
where $n=0,1,2,3...$, and defining 
\be
h_{\beta_o}(E_n)=g(E_n) e^{-\beta_o E_n},
\ee
the energy probability distribution (EPD) at the inverse temperature $\beta_o$, we can rewrite the partition function in terms of $n$, $h_{\beta_o}(E_n)$ and $\delta \beta= \beta-\beta_o$ as
\be
Z=e^{-\delta \beta E_0} \sum_n h_{\beta_o}(E_n) z^n \label{PartFunction}
\ee
with $z=e^{-\delta \beta \delta E}$. The Fisher zeros have a direct correspondence to the EPD zeros and can be determined by the (complex) roots of 
\be
P(z) = \sum_n h_{\beta_o}(E_n) z^n.
\ee

In the thermodynamic limit (infinite system) the dominant zero at the critical temperature will be $z_c=1$. For finite systems, however, the dominant root (dominant zero) is the one with the smallest imaginary part. So, the EPD zeros method consists, after determining the polynomial $P(z)$ at an initial guess critical temperature ($\beta^{(0)}_c$), in computing the roots of $P(z)$ and searching for the dominant zero $z_c$ among the zeros of this set. After that, a new approximated critical temperature can be estimated by
\be
\beta_c^{(n+1)}=\beta_c^{(n)}-\frac{\ln(Re(z))}{\delta E} \label{betac}
\ee
for $n=0$,
since the relation 
\be
Re(z_c) = e^{-(\beta_c-\beta^{(0)}_c) \delta E} \label{zc}
\ee
is expected. 
Repeating this procedure successively, a sequence of new estimative critical temperatures $\beta^{(n)}_c$ can be obtained until reaching some convergence criteria (e.g.$|\beta_c^{(n+1)}-\beta_c^{(n)}| < \epsilon$, where $\epsilon$ is a given precision).

The major computational effort in this process involves the determination of the energy distributions $g(E)$ or $h_{\beta_0}(E)$ \cite{Wolff,BroadHistogram,WangLandau1,WangLandau2}. After that, however, more computational effort is still necessary to locate the zeros of $P(z)$ and in the search of the dominant zero. Depending on the size of the system this process can become prohibitively lengthy and slow, since the order of $P(z)$ grows with the physical system dimensions.
The procedure can be accelerated, however, by selecting only the highest coefficients of $P(z)$ (neglecting those lower than certain cut-off limit, $h_{cut}$, i.e., if $\frac{h_{\beta_0}(E_n)}{h_{max}} < h_{cut}$, where $h_{max}$ is the highest $h_{\beta_0}(E_n)$). Usually, $h_{cut}=0.01$ generate good results.
However, the increase in the accuracy (lower $h_{cut}$ values) increases also the number of terms on $P(z)$, with the consequent increasing on its number of zeros. This has a great impact on the processing time of the zero search procedures, jeopardizing, in practice, the obtaining of more accurate results. The same effect (high number of terms on the polynomial $P(z)$) is expected for models with flat energy distributions, even for higher $h_{cut}$ values. Thus, to reduce the processing time in the search of the dominant zero, as well as to allow lower $h_{cut}$ values without compromising this processing time, are necessary steps to increase even more the efficacy and the accuracy of the EPD zeros method - and this is the aim of the present contribution. We will introduce, in the context of the EPD zeros method, an alternative method to find only the dominant zero of $P(z)$. As we shall show, the method increases the speed of the dominant zero search procedure, and allows the use of the entire polynomial $P(z)$ with no cut-off, increasing also the accuracy of the results.

\section{Scaled Newton-Raphson}

The Newton-Raphson method \cite{Macleod1984} is appropriated for estimative based on one single initial point. It can be obtained, in the context of the EPD zeros method, approximating the derivative of $P(z)$ at $z_0$, computed in the vicinity of a root $z_c$ of $P(z)$ by 
\be
\frac{dP(z)}{dz}\big{|}_{z_0} \equiv P'(z_0) \approx \frac{P(z_0)-P(z_c)}{z_0-z_c}.
\ee
Now, since $P(z_c)=0$, we obtain,
\be
z_c=f_P(z_0)=z_0-\frac{P(z_0)}{P'(z_0)}, \label{NewtonRaphson}
\ee
the Newton-Raphson formula. The fixed points of the transformation $f_P : z \mapsto z-\frac{P(z)}{P'(z)}$ are the roots of $P(z)$ and this transformation can be iterated and, starting from an	 initial guess point $z^{(0)}$, converges to one of the roots of $P(z)$. The key point is, thus, the choice of the initial guess $z^{(0)}$ that generates, at the ending of the process, the dominant zero of $P(z)$. This is not so trivial since, in general, starting from any arbitrary initial guess, it is more likely to the process to converge to some of the other roots than to the dominant zero.

The successive application of the following transformation 
\be
 f^{(1)}_P(z)=\frac{Re(z)}{Re\big( z-\frac{P(z)}{P'(z)} \big)} \Big( z-\frac{P(z)}{P'(z)} \Big), \label{transf1}
\ee
shows up to converge to an appropriated initial point for the Newton-Raphson procedure in order to obtain the dominant zero of $P(z)$. 
Here, $Re(\zeta)$ is the real part of the complex number $\zeta$, as $Im(\zeta)$ is its imaginary part. It should be notice that $f^{(1)}_P(z)$ preserves the real part of the function argument ($z$) and acts only in its imaginary part. So, all the complex numbers in the sequence generated by the iteration $z^{(n+1)}=f^{(1)}_P(z^{(n)})$ starting from an initial guess $z^{(0)}=1+i\delta$ have their real parts equal to $1$.  

The fixed point of $f^{(1)}_P(z)$ is the solution of
\be
z=Re(z)+i \frac{Im(z)-Im(\frac{P(z)}{P'(z)})}{Re(z)-Re(\frac{P(z)}{P'(z)})},
\ee
and, as $z=Re(z) + i Im(z)$, we obtain that 
\be
Re(z) Im\big(\frac{P(z)}{P'(z)}\big)- Im(z) Re\big(\frac{P(z)}{P'(z)}\big)=0
\ee
at the fixed point of $f^{(1)}_P(z)$.

The following procedure can thus be applied in order to obtain the dominant zero of the polynomial $P(z)$: (i) starting from an initial point $z_0=1+i\delta$ we successively apply the transformation $f^{(1)}_P(z)$ until reach a given precision 
\be
|Re(z) Im\big(\frac{P(z)}{P'(z)}\big)- Im(z) Re\big(\frac{P(z)}{P'(z)}\big)| \leq \epsilon_1,
\ee
obtaining an appropriate initial point $z^{(0)}$ for the Newton-Raphson procedure; (ii) we then use the initial point $z^{(0)}$ to iterate the application of $f_P(z)$ up to the desired precision $|P(z)| \leq \epsilon_2$. From now on we will refer to this procedure as the scaled Newton-Raphson (SNR) method.

\begin{figure}[ht]
\begin{center}
\includegraphics[angle=0,width=0.5\textwidth]{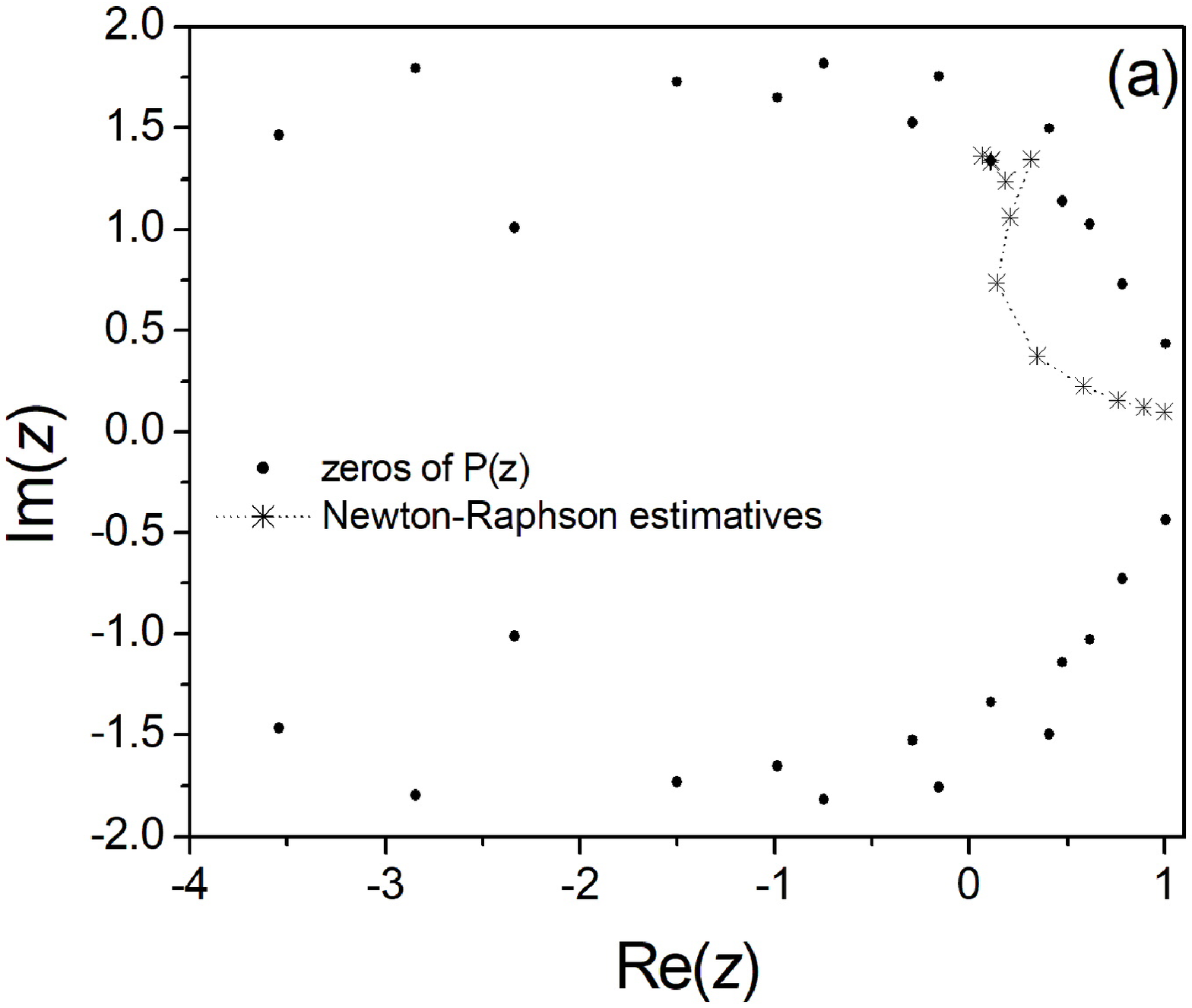}\includegraphics[angle=0,width=0.5\textwidth]{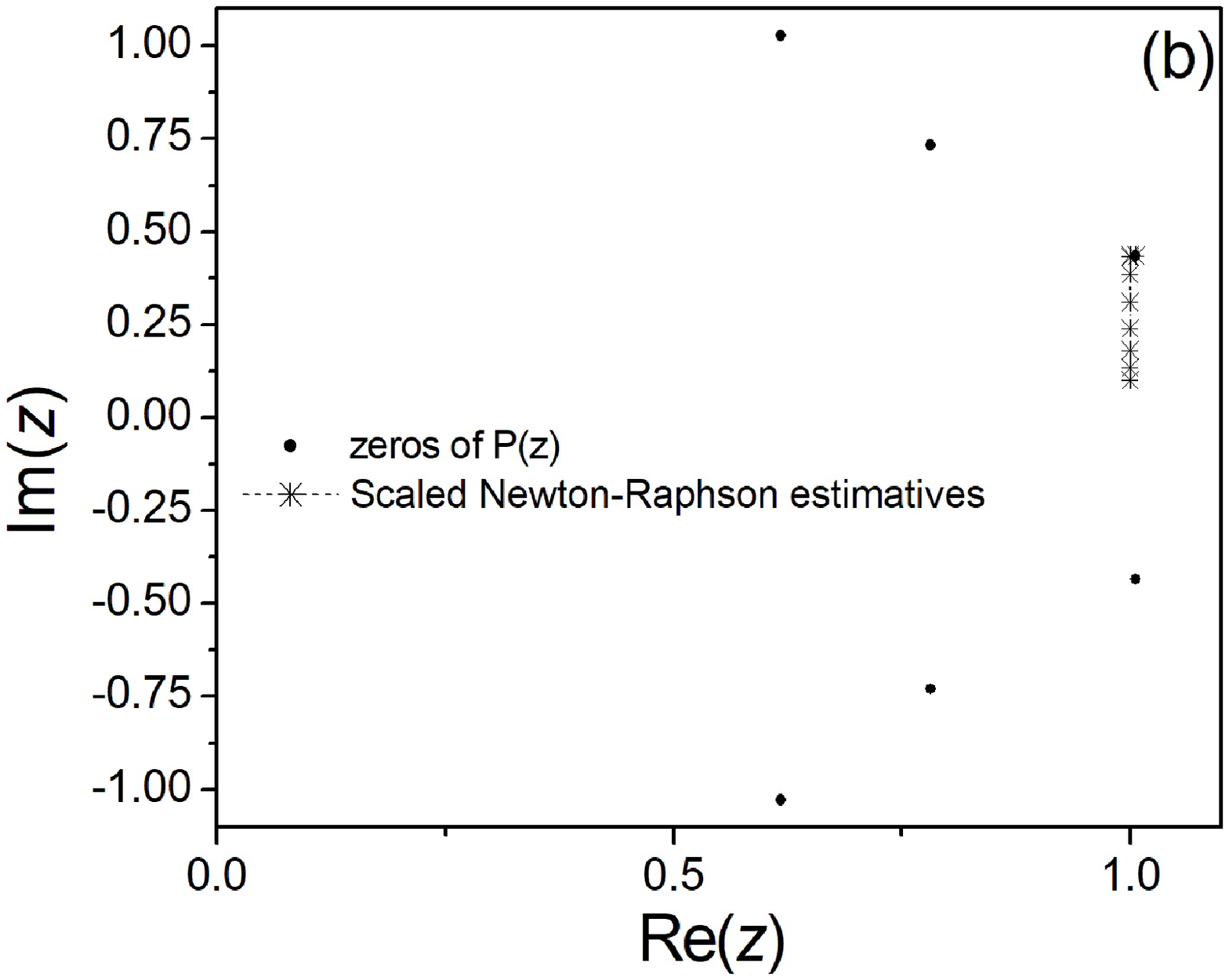}
\end{center}
\caption{Trajectory of the estimative for the dominant zero for the Newton-Raphson procedure (a) and with the SNR method (b). The dots are the Fisher zeros for the polynomial given by eq. (\ref{Pz8x8}). The asterisks are the Newton-Raphson or SNR estimative, starting from the same initial point. Newton-Raphson procedure deviates from the dominant zero and converges to another Fisher's zero. SNR converges, after few steps, to the dominant zero.}
\label{FigPath}
\end{figure}

In order to illustrate the SNR method, we show, in Fig. \ref{FigPath} the zeros of the following polynomial of 64th order:
\begin{eqnarray}
P(z) &=& .323646\times 10^{-48}z^{64}+.661146\times 10^{-45}z^{62}+.747054\times 10^{-44}z^{61} \nonumber \\
&& +.770264\times 10^{-42}z^{60}+.166916\times 10^{-40}z^{59}+.765197\times 10^{-39}z^{58} \nonumber \\
&& +.211971\times 10^{-37}z^{57}+.718005\times 10^{-36}z^{56}+.209995\times 10^{-34}z^{55} \nonumber \\
&& +.635611\times 10^{-33}z^{54}+.184979\times 10^{-31}z^{53}+.539143\times 10^{-30}z^{52} \nonumber \\
&& +.155114\times 10^{-28}z^{51}+.441074\times 10^{-27}z^{50}+.122491\times 10^{-25}z^{49} \nonumber \\
&& +.329050\times 10^{-24}z^{48}+.845373\times 10^{-23}z^{47}+.205729\times 10^{-21}z^{46} \nonumber \\
&& +.470163\times 10^{-20}z^{45}+.100191\times 10^{-18}z^{44}+.197856\times 10^{-17}z^{43} \nonumber \\
&& +.360078\times 10^{-16}z^{42}+.600683\times 10^{-15}z^{41}+.913627\times 10^{-14}z^{40} \nonumber \\
&& +.125998\times 10^{-12}z^{39}+.156666\times 10^{-11}z^{38}+.174647\times 10^{-10}z^{37} \nonumber \\
&& +.173608\times 10^{-9}z^{36}+.153124\times 10^{-8}z^{35}+.119324\times 10^{-7}z^{34} \nonumber \\
&& +.818780\times 10^{-7}z^{33}+.493673\times 10^{-6}z^{32}+.261346\times 10^{-5}z^{31} \nonumber \\
&& +.121570\times 10^{-4}z^{30}+.497954\times 10^{-4}z^{29}+.180204\times 10^{-3}z^{28} \nonumber \\
&& +.578637\times 10^{-3}z^{27}+.165679\times 10^{-2}z^{26}+.425310\times 10^{-2}z^{25} \nonumber \\
&& +.984367\times 10^{-2}z^{24}+.206578\times 10^{-1}z^{23}+.395259\times 10^{-1}z^{22} \nonumber \\
&& +.693242\times 10^{-1}z^{21}  +.112051z^{20}+.167835z^{19}+.234410z^{18} \nonumber \\ 
&& +.307453z^{17}+.381980z^{16}+.453867z^{15}+.521661z^{14}  \nonumber \\
&& +.585565z^{13}+.649647z^{12}+.711449z^{11}+.780301z^{10} \nonumber \\
&& +.822862z^{9}+.898041z^{8}+.846233z^{7}+.975076z^{6} \nonumber \\
&& +.678909z^{5}+1.00000z^{4}+.309572z^{3}+.874496z^{2}+.436141 \label{Pz8x8}
\end{eqnarray}

Starting from the same initial point, $z=1+0.1i$, the successive iterations of the Newton-Raphson procedure deviate from the dominant zero and converge, after 14 iterations, to a different zero ($0.113815+1.33883$, fig.\ref{FigPath}.a) whereas the SNR converges, after only 11 iterations, to the actual dominant zero ($1.0061+0.43561i$, fig.\ref{FigPath}.b).

We will compare the processing time of the SNR procedure with the search procedure employed on the EPD zeros method \cite{Rocha2016, bvcosta2017, bvcosta20172, Figueiredo2017, Mol2018, bvcosta2019, bvcosta20192, Mol2019}. In these previous works, the zeros of the EPD polynomials are computed employing the built-in procedures of algebraic processors and then the list of zeros is scanned to search for the one closest to the thermodynamic limit dominant zero ($z=1$). Other procedures to locate the partition function zeros were also employed before on the study of phase transitions on discrete spin models, as, for example, the work from Alves, Felicio and Hansmann \cite{Alves1997}. Their approach is a two step procedure, first performing a (graphical) search of a common zero for the real and imaginary parts of the partition function, followed by an unconstrained nonlinear minimization procedure. In this sense, this procedure involves also a search step, as the one employed on the EPD zeros method.

\begin{figure}[ht]
\begin{center}
\includegraphics[angle=0,width=0.5\textwidth]{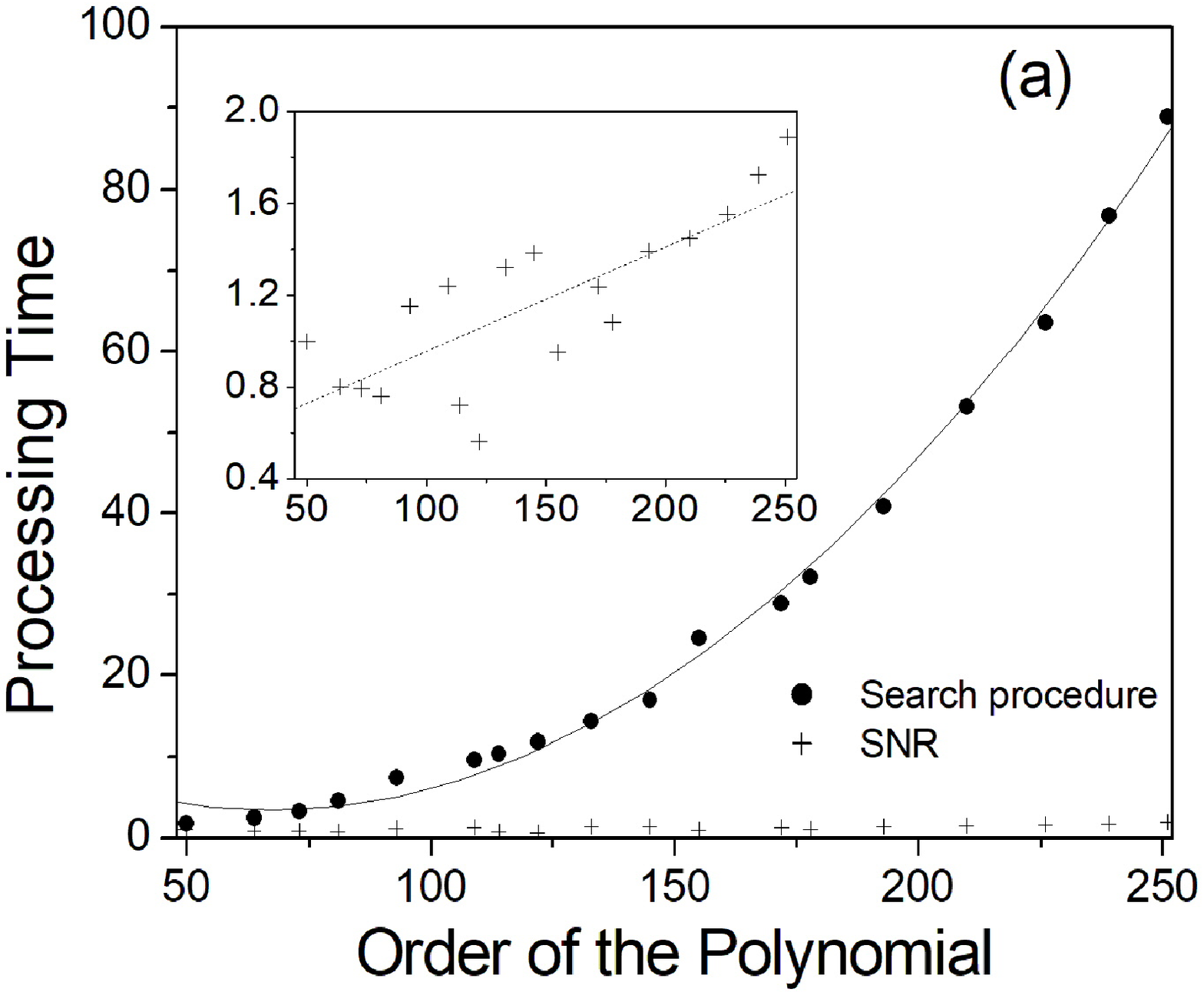}\includegraphics[angle=0,width=0.5\textwidth]{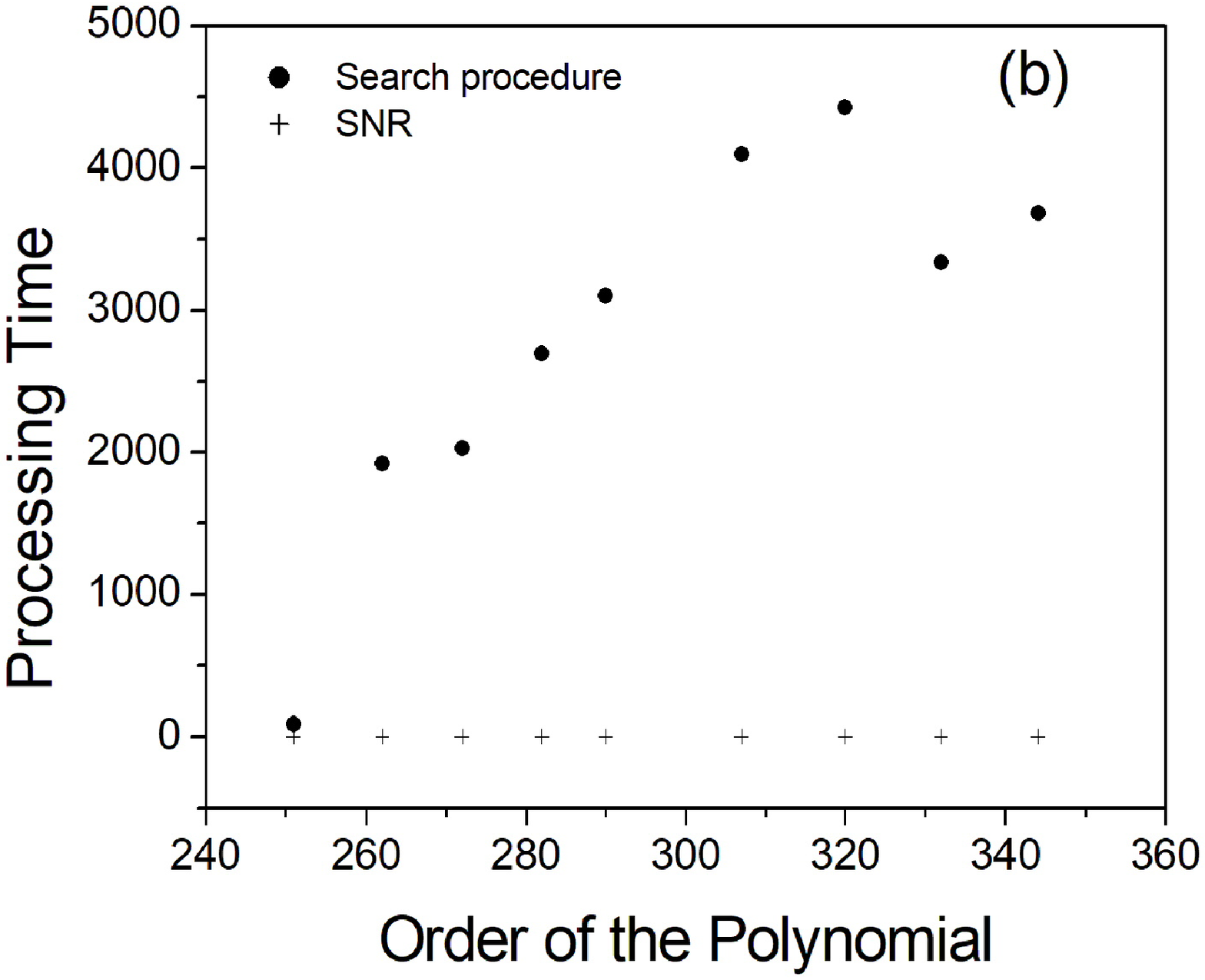}
\end{center}
\caption{Normalized processing time for locating the dominant zeros by using the search procedure (full dots) and the SNR procedures (crosses) of polynomials of different orders. (a) Up to a polynomial of order $250$ the processing time for the search procedure grows quadratically (least squares fit represented by the solid line), while the processing time for the SNR procedure is about two orders of magnitude lower and shows an approximately linear growth. (b) For a polynomial order greater than $250$, the processing time to locate the dominant zero in the search procedure strongly grows in order of magnitude, while the smooth growth of the SNR procedure is maintained.}
\label{FigTime}
\end{figure}

In the present work, we implemented the partition function zeros search procedure in the algebraic processor {\it{Maple\textsuperscript{\textregistered}}}, by using the built-in procedure {\it fsolve}. The SNR procedure was implemented as an interpreted sequence of commands on the same environment. The comparison between the processing times spent by using the search procedure and the SNR procedure can be found on both Table \ref{tab1} and figure \ref{FigTime}. In Table \ref{tab1} we present, in the third column, the processing time ratio between the SNR and the search procedure for several 2D Ising lattice sizes, as we will discuss in section 3. The SNR method is always faster than the search procedure, and for a $200 \times 200$ lattice, the SNR procedure is a thousand times faster than the search procedure. In Fig. \ref{FigTime}, we compare the processing time (averaged over different runs) necessary to locate the dominant zero of several polynomials of orders running from $50$ to $350$. The processing time is normalized in a scale where the time spent to locate the dominant zero of a polynomial of order $50$ employing the SNR procedure is set to $1$.  We observe, in fig. \ref{FigTime}(a), that while the computer processing time grows quadratically with the polynomial order for the search procedure (full dots interpolated by the continuous line), the SNR procedure presents only a smooth linear growth (crosses, and dotted line on the inset) and it is two orders of magnitude lower than the search procedure processing time. For polynomials of order greater than $250$, the computer processing time spent on the search procedure is even higher, becoming even one thousand times slower than the SNR procedure (fig. \ref{FigTime}(b)). 

\section{Application to the 2D Ising Model}

To exemplify the application of the method, we apply it to the 2D Ising model \cite{Onsager1944} on a $L \times L$ square lattice, whose Hamiltonian is given by
\be
H = -J\sum_{<ij>} S_i S_j,
\ee 
where $i$ and $j$ are the sites of the lattice, $J$ is the exchange energy (with $J>0$, i.e., ferromagnetic coupling) and $S_i= \pm 1$ are the Ising spin variables. In what follows we use $J=1$, i.e., the quantities are described in units of $J$.

The exact probability distribution, $g(E)$, of the two-dimensional Ising model on a square finite lattice is determined by using the Beale procedure \cite{Beale1996} obtained from the Kaufman's generalization \cite{Kaufman1949} of the Onsager's solution \cite{Onsager1944}. The polynomial presented on eq.(\ref{Pz8x8}) was obtained from the partition function for an $8 \times 8$ square lattice at $T=2.31$, with a 20 digits precision (not all digits shown in eq. (\ref{Pz8x8})). 

\begin{table}
\centering
\caption{Critical temperature, processing time ratio (see text), and the imaginary part of the dominant zero for the 2D Ising model within the EPD zeros method with SNR procedure for different cut-offs and lattice sizes.} 
\begin{tabular}{ccccc}
\hline
  Lattice   &   $h_{cut}=0.01$ & $h_{cut}=0.01$ & $h_{cut}=0$ & $h_{cut}=0$ \\
  size (L)  &  $T_c$ & time ratio & $T_c$ & $Im(z_c)$ \\
\hline
  12 & $2.31907$ & $0.1021^a$ & $2.31819$ & $0.28223$ \\
  16 & $2.31292$ & $0.3327$E-2$^a$ & $2.31192$ & $0.21039$\\
  20 & $2.30714$ & $0.2044$E-3$^a$ & $2.30622$ & $0.16797$ \\
  24 & $2.30252$ & $0.1921^b$ & $2.30161$ & $0.13987$\\
  28 & $2.29877$ & $0.1436^b$ & $2.29794$ & $0.11987$\\
  32 & $2.29557$ & $0.1374^b$ & $2.29497$ & $0.10489$\\
  36 & $2.29317$ & $0.0367^b$ & $2.29253$ & $0.09325$\\
  64 & $2.28472$ & $0.0146^b$ & $2.28310$ & $0.05257$\\
 200 & $2.31009$ & $ < 1 \times 10^{-3}$ & $2.29179$ & $0.02013$\\
\hline
\end{tabular}
\label{tab1}
\end{table} 

If we start from an initial temperature $T=2.4$ and from the point $z_0=1+\delta i$ with $\delta=0.1$, the convergence of $f^{(1)}_P(z)$ is achieved after 11 iterations with precision $\epsilon_1=10^{-2}$, generating the initial point $z^{(0)}=1+0.4396791976i$. We verified that any value of $0.05 \leq \delta \leq 0.5$ results in the same initial point. Then, applying the Newton-Raphson method, after only 5 iterations the procedure converges with precision $\epsilon_2=10^{-6}$ to $z=0.9428476996+0.4082250676i$, the estimated dominant zero. It differs from the exact dominant zero computed with the {\it{solve}} built-in function of the algebraic processor {\it{Maple\textsuperscript{\textregistered}}} only at the 9th significant figure. The mean ratio between the execution time for the {\it{solve}} built-in procedure (followed by the search of the dominant zero) and the execution time for the SNR method, computed in the same environment (hardware+software), is, in this case, $0.0103$. Then we apply Eq.(\ref{betac}) to compute the estimated critical temperature obtaining $T_c \approx 2.318145563$. It is worth to mention that this first approximated result for $T_c$ was obtained by employing all terms of the partition function (without cut-off) and, for this reason, it is only $0.4\times 10^{-7}\%$ greater than the final exact value. If we employ a cut-off $h_{cut}=0.01$, the first estimative for the critical temperature will result in $T_c=2.318536818$ ($0.017\%$ greater than the exact result) and, after convergence, $T_c=2.318125068$ ($8 \times 10^{-4}\%$ lower), still very accurate results.

\begin{figure}[ht]
\begin{center}
\includegraphics[angle=0,width=0.9\textwidth]{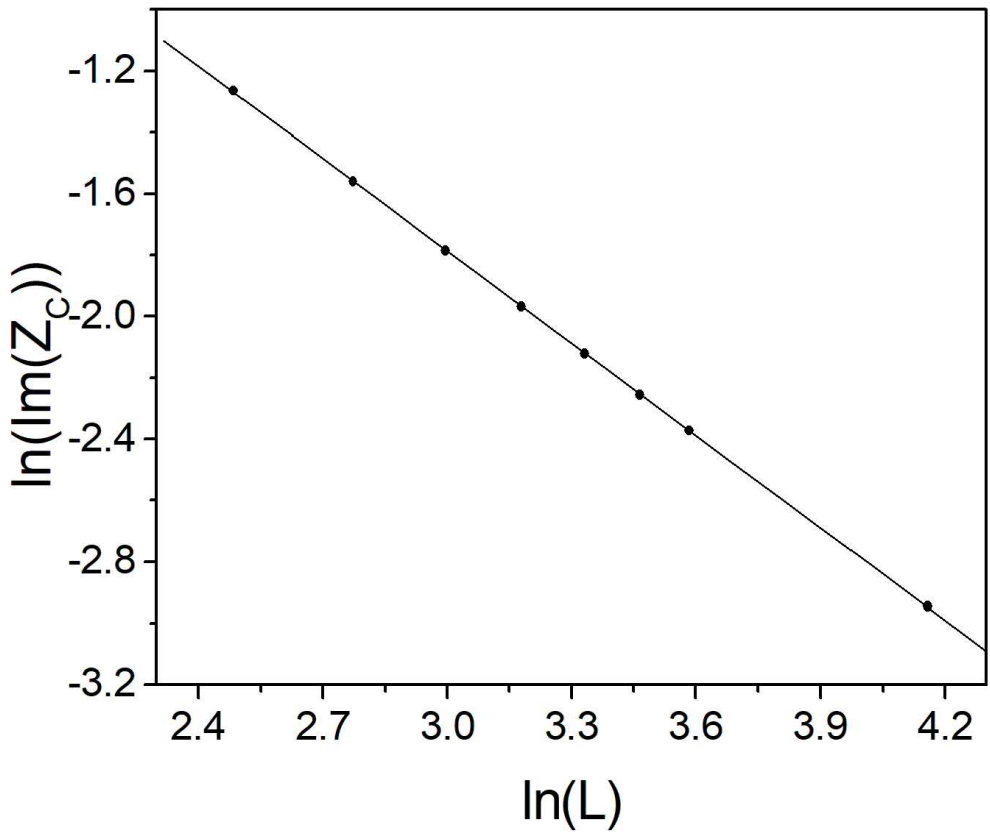}
\end{center}
\caption{Finite size scaling of $Im(z_c)$ (filled circles) and its linear fit (solid line), estimating the critical exponent $1/\nu=1.00(3)$.}
\label{LnzLnL}
\end{figure}

Results for different lattice sizes are presented on Table \ref{tab1}. We computed the critical temperatures for a set of $L \times L$ square lattices with $L=12,16,20,24,28,32,36$ and $64$ (EPD obtained from the Beale procedure\cite{Beale1996}) and $L=200$ (EPD obtained from the Wang-Landau method \cite{WangLandau1,WangLandau2}). For $L \le 36$ the EPD was obtained with infinite precision, by using the big integers capabilities of the algebraic processor employed, setting the number of digits of precision to the first integer greater or equal to $L^2 \log_{10}(2)$\cite{Beale1996}. The $L=64$ results were computed with float pointing approximations. The details of the Wang-Landau procedure employed in the obtaining of the $L=200$ result will be presented latter, the error on $T_c$ is greater, in this case, due to the errors on the EPD. 

Differences between the results computed with a cut-off $h_{cut}=0.01$ (second column of Table \ref{tab1}) and without cut-off, i.e., $h_{cut}=0$ (fourth column) are only in the 4th significant figure. The ratios between the mean processing time for the implemented SNR method and the mean processing time spent employing the {\it{Maple\textsuperscript{\textregistered}}} algebraic processor built-in procedures {\it solve} or {\it fsolve} (averaged over successive repetitions of the procedures), for $h_{cut}=0.01$, are displayed on the third column. For a lattice size $L = 24$ approximately the processing time for the {\it solve} procedure became prohibitive, and we used, from this point on, the numerical {\it fsolve} procedure. In both cases the implemented SNR procedure is faster than the other methods, even tough that, in the environment it was implemented ({\it{Maple\textsuperscript{\textregistered}}}), SNR is an interpreted routine and the {\it solve} or {\it fsolve} are internal compiled routines. For $h_{cut}=0$ the execution time for both {\it solve} or {\it fsolve} became prohibitively high, but the SNR method remains fast.

\begin{figure}[ht]
\begin{center}
\includegraphics[angle=0,width=0.9\textwidth]{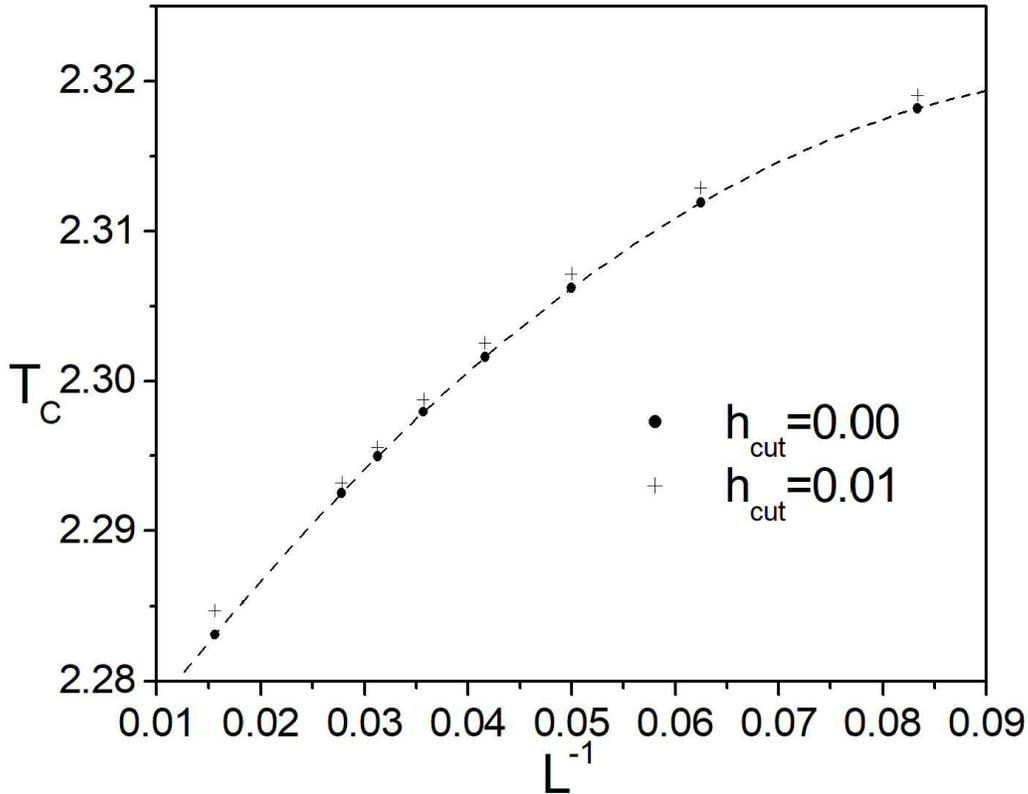}
\end{center}
\caption{The critical temperature $T_c$ (dots) for $h_{cut}=0$ as a function of the inverse lattice size $1/L$. The critical temperature exhibits a quadratic behavior with $L^{-1}$. In the thermodynamic limit the fit to a second order polynomial predicts $T_c=2.2689(8)$, to be compared to the exact result $T_c=2.26919$. The crosses represent the critical temperatures computed with $h_{cut}=0.01$.}
\label{TcxL_1}
\end{figure}

\begin{figure}[ht]
\begin{center}
\includegraphics[angle=0,width=0.9\textwidth]{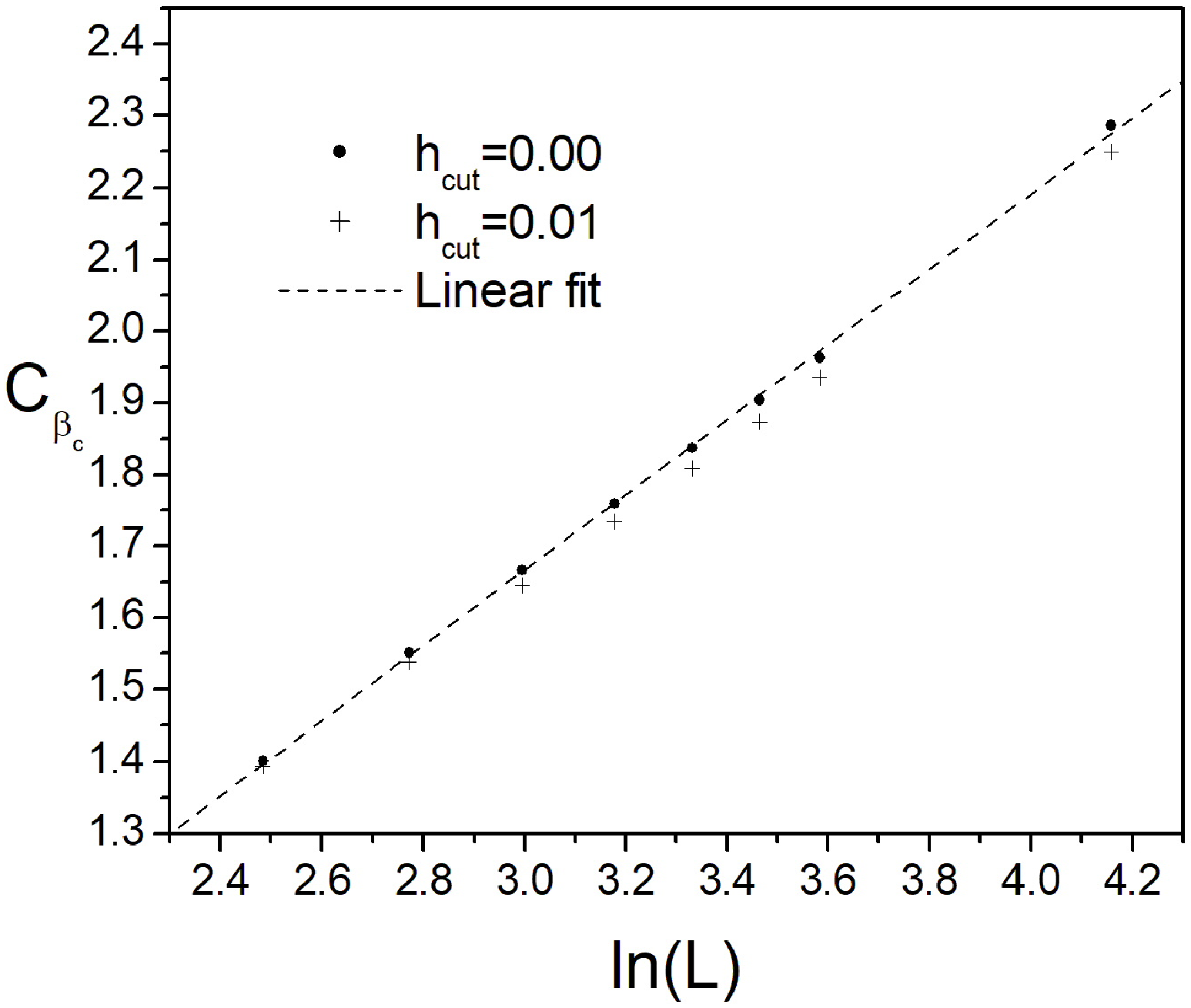}
\end{center}
\caption{The specific heat per spin at the critical temperature ($c_{\beta_c})$ as a function of $\ln(L)$, for $h_{cut}=0.00$ (dots) and $h_{cut}=0.01$ (crosses). Both set of data fit well the predicted relation $c_{\beta_c} = A \ln(L) + B$. If we include finite size corrections, $c_{\beta_c} = A \ln(L) + B + C L^{-1} + D L^{-2}$, the fitting to the calculated data results in an accurate result for $h_{cut}=0.00$ when compared to the Onsager's expected result, whereas for $h_{cut}=0.01$ the fitting results deviates from the expected ones (see text for details).}
\label{cbetaclnl}
\end{figure}

The gain on accuracy is exemplified on Fig. \ref{LnzLnL}, that brings the linear fit for the log-log plot of the imaginary part of the dominant zero (for $h_{cut}=0$). Results are computed from the EPD obtained by the Beale procedure, i.e., the results from Table \ref{tab1} for $L \le 64$. The linear fit to the finite size scaling theory expected behavior for the 2D Ising model\cite{FiniteSizeIsing2D}, $Im(z_c(L)) \sim Im(z_c) + b L^{-1/\nu}$ \cite{bvcosta2017} gives, for the critical exponent $\nu$, the estimative value $1/\nu = 1.00(3)$, very close to the the exact value $\nu=1$, with a Pearson correlation coefficient $r=0.9999(9)$. Fig. \ref{TcxL_1} shows the $2^{nd}$ order polynomial fit (solid line) of the critical temperature $T_c$ as a function of $L^{-1}$ for $h_{cut}=0$ (full dots). The results for $h_{cut}=0.01$ are also displayed (crosses) in order to illustrate the correction represented by employing all the terms of $P(z)$ on the computation of the zeros. The polynomial fit, $T_c(L) = T_c +b L^{-1} + c L^{-2}$, is compatible with the finite size scaling theory since, for large $L$, the expected behavior is $T_c(L) \sim T_c +bL^{-1}$. The squared Pearson correlation coefficient (the determination coefficient), in this case, is $r^2=1.0000(0)$, and the critical temperature in the thermodynamic limit ($L \rightarrow \infty$) is estimated as $T_c=2.2689(8)$, also very close to the exact Onsager result, $T_c=2.26919$ \cite{Onsager1944}. 

\begin{table}
\centering
\caption{Heat capacity at critical temperature for different lattice sizes with and without cut-offs.}
\begin{tabular}{ccc}
\hline
  Lattice   &   $h_{cut}=0.01$ & $h_{cut}=0.00$  \\
  size (L)  &  $c_{\beta_c}$ & $c_{\beta_c}$  \\
\hline
 12 & 1.393256 & 1.401167 \\
 16 & 1.537419 & 1.551459 \\
 20 & 1.645465 & 1.666132 \\
 24 & 1.734387 & 1.758980 \\
 28 & 1.808398 & 1.837039 \\
 32 & 1.873923 & 1.904397 \\
 36 & 1.934800 & 1.963647 \\
 64 & 2.248601 & 2.286275 \\
200 & 1.671064 & 2.022945 
\\\hline
\end{tabular}
\label{tab2}
\end{table} 

We also evaluated the behavior of the specific heat per spin at the transition temperature, $c_{\beta_c}$ as a function of the lattice size. The specific heat per site is given by
\be
c=\frac{\beta^2}{L^2}(<E^2>-<E>^2),
\ee
where, given the energy probability distribution $h_{\beta}(E)$, the mean energy $<E>$ and mean square energy $<E^2>$ at inverse temperature $\beta$ can be easily computed from 
\be
<E^k> = \frac{e^{-\delta \beta E_0}}{Z}  \sum_n (E_n)^k h_{\beta}(E_n) z^n,
\ee
with $Z$ given by eq. (\ref{PartFunction}). Within the finite size scaling theory, at a temperature near $T_c$, the specific heat scales with $\alpha/\nu$, where $\alpha$ is the specific heat exponent. For the 2D Ising model we have $\alpha = 0$, so no fitting between $\ln(c_{\beta_c}) \times \ln(L)$ is expected. However, the critical heat capacity shows a logarithmic dependence on the dimension\cite{Onsager1944, Rosengren2004, Kryzhanovskya2018} expressed by $c_{\beta_c} \approx 0.494539 \ln(L) + 0.187903$, the behavior expected for large $L$. In Fig. \ref{cbetaclnl} we show our results for the specific heat at the critical temperature (displayed on Table \ref{tab2}) computed with $h_{cut}=0$, as a function of $\ln(L)$ (full dots and solid line) and with $h_{cut}=0.01$ (crosses). Differences between the two set of data are at maximum of $1.6\%$, and both set fit well the relation $c_{\beta_c} = A \ln(L) + B$. However, one should expect some finite size corrections to this relation for small/medium size lattices \cite{Ferdinand1969}, as we observed in the temperature results, Fig. \ref{TcxL_1}. Assuming a finite size relation $c_{\beta_c} = A \ln(L) + B + C L^{-1} + D L^{-2}$ we get, for $h_{cut}=0$, $A=0.49522$ and $B=0.19529$, with a Pearson determination coefficient $r^2=1.00000(0)$, whereas, for $h_{cut}=0.01$, $A=0.592771$ and $B=-0.29168$, with $r^2=0.99999(7)$. The discrepancy of the $h_{cut}=0.01$ results to the Onsager expected values, specially for the $B$ coefficient, illustrates the importance of getting accurate results when we are dealing with small/medium lattices - if the numerical/truncation errors are of the same order of the finite size corrections, one cannot correctly preview the parameters behavior as a function of the system dimension, jeopardizing the obtaining of the results on the $L \rightarrow \infty$ thermodynamic limit. 

These two results - the $T_c$ estimative in the thermodynamic limit and the coefficients of the divergent behavior of the heat capacity at critical temperature $c_{\beta_c}$ computed with $h_{cut}=0$ illustrate the gain provided by the accuracy of the EPD zeros method together with the SNR procedure: it made possible to obtain very accurate results from small/medium lattices, reducing the errors of estimative based on those not so large lattices. It is worth to remark that more accurate EPDs are expected exactly in small/medium lattice cases since it implies a smaller number of available states, and then sampling procedures can generate more accurate EPDs.

\section{Larger Lattices}

In order to evaluate the applicability and efficacy of the SNR procedure for larger lattices (or larger EPDs) we sampled the EPD of the 2D Ising model in a $L \times L$ square lattice with $L=200$ by using the Wang-Landau method \cite{WangLandau1,WangLandau2}. The EPD was computed by performing multiple random walks on $16$ parallelized ranges of energy, with two convergence criteria: modification factor $f \le 1\times 10^{-8}$ and histogram flatness greater than $93\%$. We then use the resulting energy probability distribution in the EPD zeros method employing both the search procedure and the SNR procedure for a cut-off $h_{cut}=0.01$, in order to compare the processing time, and the SNR procedure with no cut-off, in order to generate more accurate results. As we can see, from the last line of Table \ref{tab1}, the results are not so accurate as the ones computed with the Beale procedure (for $L \le 64$) - we can estimate the expected exact values for a $L \times L$ lattice, with $L=200$, from the scale fits obtained from the small or medium size lattices, i.e., $T_c=2.2737$, $Im(z_c)=0.01670$ and $C_{\beta_c}=2.81813$. However, this loss in accuracy is expected, since the results for $L \le 64$ where generated by using the exact energy probability distribution, whereas the $L=200$ results use the sampled EPD from the Wang-Landau method. The main result to be highlighted here is the improvement in both processing time and accuracy when the SNR method is applied. As already mentioned, the computer processing time in the SNR case is more than one thousand times faster than the search procedure, for a cut-off $h_{cut} = 0.01$. The SNR procedure also allows the computation of the dominant zero even when no cut-off is applied, corresponding to a polynomial of order $40001$, with estimative results for $T_c$, $Im(z_c)$ and $C_{\beta_c}$ closer to the expected results (numerical values on Tables \ref{tab1} and \ref{tab2}) than those for $h_{cut} = 0.01$. In addition, the processing time to locate all the zeros within the search procedure was shown to be prohibitive.

\section{Conclusion}

In summary, we introduced the scaled Newton-Raphson method to locate the dominant zero within the EPD zeros method context. We have shown that this procedure strongly reduces the execution time, not only speeding up the computer processing of the EPD zero method but also allowing its application with higher accuracy. Besides, the SNR method can be easily implemented in compiled computer languages, allowing the implementation in one single computer code of both the energy probability distribution evaluation and the EPD zeros method. For the 2D Ising model, we have shown the gain in the computer processing time and accuracy when SNR is applied together with the EPD zeros method. Due to the increased accuracy provided by the use of the SNR procedure with $h_{cut}=0$, even using the results from small/medium lattices, it was possible to obtain very accurate estimative for thermodynamic quantities (critical temperature, critical exponent, and heat capacity at critical temperature) in the thermodynamic limit.



\section*{Acknowledgments}

This research was supported by FAPEMIG and CAPES (Brazilian funding agencies).

\end{document}